\documentclass[10pt,a4paper]{article}
\usepackage[T1]{fontenc}
\usepackage[utf8]{inputenc}
\usepackage{lmodern,microtype,amsmath,amssymb,booktabs,tabularx,longtable,array,multirow,graphicx,caption,xcolor,enumitem}
\usepackage[margin=0.88in,top=0.92in,bottom=0.87in]{geometry}
\usepackage[round,authoryear]{natbib}
\usepackage[colorlinks=true,urlcolor=blue!65!black,citecolor=blue!65!black,linkcolor=blue!60!black]{hyperref}
\usepackage{fancyhdr}
\definecolor{aiso}{HTML}{184AEB}\definecolor{ink}{HTML}{1B263B}
\setlist{nosep,leftmargin=*,topsep=3pt}
\graphicspath{{figures/}}
\newcommand{\pnt}[1]{\textbf{#1}}

\title{\vspace{-1.2cm}\Huge\bfseries ChatGPT Can't Sell\\[2pt]\Large\bfseries (And That's Probably a Good Thing)\\[12pt]\large Buyer-Aligned AI Advice, the Sales-Representative Gap, and Evidence from Commercial Conversations}
\author{\textbf{Benjamin Tannenbaum}\\\small Aiso Research, AISO Boost Ltd.\\\small \url{https://getaiso.com/}}
\date{21 November 2026\\\small Research preprint, version 2.0 | Prepared for arXiv submission}
\hypersetup{pdftitle={ChatGPT Can't Sell (And That's Probably a Good Thing): Buyer-Aligned AI Advice and the Sales-Representative Gap},pdfauthor={Benjamin Tannenbaum, Aiso Research}}
\begin{document}
\maketitle
\begin{abstract}
AI assistants and merchant sales agents have different objectives: assistants should help buyers compare options and preserve the choice not to purchase, while sales agents also guide qualified buyers toward an authorized next step. We study this sales-representative gap with three evidence streams. In 12,178 historical English GPT-family conversations screened for commercial or transactional intent, a 279-request sample contained recommendation language in 63 first replies (22.6\%) and no narrow transactional action wording. In a 56-conversation explicit-buying sample, 24 replies (42.9\%) contained recommendation markers, 32 (57.1\%) qualification phrases, and none transaction-specific phrases. These lexical measures do not estimate successful selling. A 2026 brand-monitoring cohort of 1,050 GPT-5.4--Gemini answer pairs over 383 purchase-related prompts showed transaction-specific phrases in 0.0\% of recommendation-worded prompts, 0.21\% of price/comparison prompts, and 9.52\% of explicitly transaction-worded prompts. The prompts are selected, not organic shopping sessions. In a prospective pilot across five matched hypothetical product scenarios, sales instructions produced more next-step language than general-assistant instructions (5/5 vs. 2/5), but no increase in clean product choice (3/5 in both); unsupported offers and fit errors also occurred. This pilot cannot establish a population effect. A separate company-affiliated randomized field study reports higher engaged-visitor conversion for Javo, a commercially trained sales agent, than for a frontier-model agent in shared tooling, but does not identify the mechanism. We argue that apparent reluctance to sell can be appropriate for buyer-aligned assistants. Merchant agents should optimize qualified decision progress through catalog grounding, truthful authorization, and safeguards against unsuitable sales.
\end{abstract}

\section{Introduction}
\emph{ChatGPT can't sell} is an intentionally provocative headline. It does not mean that ChatGPT cannot recommend a product, describe a price, or occasionally make a persuasive case. It names a narrower organizational mismatch: a general-purpose assistant is expected to help the \emph{user} reach an informed decision, whereas a brand's on-site representative is expected to help the \emph{merchant} achieve a legitimate commercial outcome. These goals often align when the product fits. They can diverge when the product does not fit, a term is unfavorable, a cost is uncertain, or an agent invents an action it is not empowered to execute.

A comparison of two laptops may be excellent advice while failing every standard sales-workflow metric. A conversion-maximizing recommendation may increase an order count yet mislead the buyer. Neither follows simply from the number of hedges, disclaimers, or affirmative phrases used by an LLM. The question is whether an assistant recognizes the boundary between advice and selling and whether a specialist can bridge that boundary without compromising consumer welfare.

This tension is timely. OpenAI's published Model Spec emphasizes honesty, user agency, respect for instructions, and the control of side effects \citep{openai2026spec,openai2026approach}. The persuasion-knowledge literature predicts that buyers notice selling intentions and may react when they infer self-interested recommendations \citep{friestad1994persuasion,campbell2000persuasion}. Dialogue-based post-training can make models broadly more useful while introducing style or agreeableness incentives that deserve separate evaluation \citep{ouyang2022training,sharma2024sycophancy}. Experimental research on AI sellers also finds that salient profit motives change consumer reactions \citep{farmer2026profit,liu2025manipulate}. A user-oriented assistant's reluctance to close a sale can therefore be socially useful rather than a defect.

For brands, however, information alone may not turn a qualified inquiry into a completed, authorized transaction. Sales competence requires preference elicitation, fit assessment, objection handling, selective recommendation, and a credible next step \citep{su2026sell,wang2026b2b}. The appropriate commercial question is therefore not \emph{how do we make ChatGPT less careful?} but \emph{how do we build an agent that can sell without violating the buyer's interests?}

This paper makes four contributions. First, it offers a first-response audit of direct purchase requests from an Aiso conversation dataset rather than relying exclusively on invented buyer scenarios. Second, it cross-checks the tendency toward informational versus transaction-specific language in recently recorded GPT responses. Third, it reports a controlled model-role experiment that illustrates the trade-off between prompting for commercial initiative and maintaining product fit and action grounding. Fourth, it positions these observations against independently reported outcomes from Javo, a sales model trained on commercial dialogues, while separating company-reported field evidence from our new mechanisms and observations.

\begin{figure}[tb]
\centering\includegraphics[width=.94\linewidth]{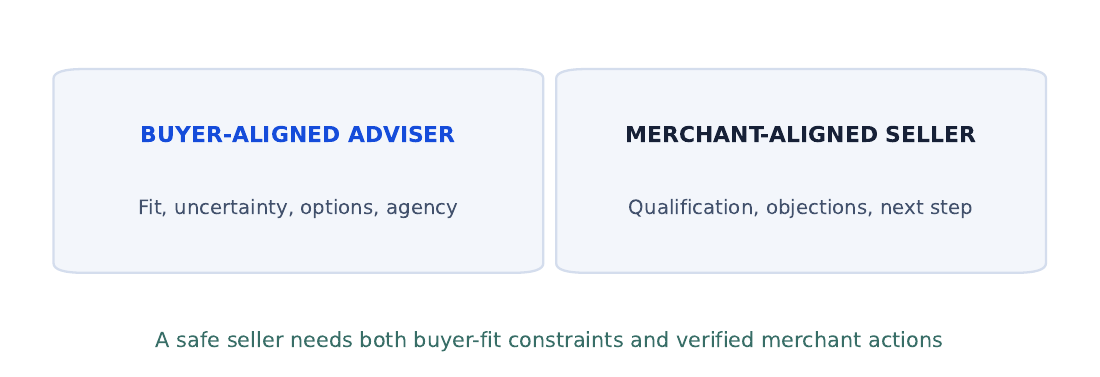}
\caption{The structural distinction motivating the paper. A trustworthy merchant agent needs the buyer-focused qualities of an adviser and the execution capabilities of a seller. This is a conceptual diagram, not an empirical causal pathway.}
\label{fig:objectives}
\end{figure}

\section{Related literature and hypotheses}
\subsection{Buyer agency, persuasion, and trust}
The Persuasion Knowledge Model distinguishes the target's beliefs about an influence attempt from the claims contained in a message \citep{friestad1994persuasion}. Consumers are more likely to scrutinize a recommendation when the commercial motive becomes salient or accessible \citep{campbell2000persuasion}. Recent AI-specific work considers whether AI-assisted selling triggers similar resistance \citep{liu2025manipulate}; \citet{farmer2026profit} report that explicit profit-motive signals can reduce the willingness to pay for AI-recommended products relative to human recommendations. These findings are relevant to \emph{why} a general-purpose adviser might avoid acting as a commissioned seller. They do not prove that a model's apparent caution was caused by post-training policies.

Related work on algorithm aversion and algorithm appreciation establishes that trust in algorithmic recommendations is conditional and context-dependent \citep{dietvorst2015aversion,logg2019appreciation}. A qualified answer is sometimes the correct one, particularly when the user's requirements are incomplete, a product's accessibility or compatibility is unverified, or the agent lacks checkout authorization. Conversely, a context-complete buyer can rationally prefer a concise answer with a transparent trade-off.

Earlier field research on sales-chatbot disclosure illustrates that users can react sharply to the perceived identity and purpose of a sales agent, a reason to separate seller claims from buyer-welfare outcomes \citep{luo2019machines}.

\subsection{Commercial dialogue is not ordinary question answering}
Recent sales benchmarks explicitly model the seller's commercial goals and the buyer's resistance \citep{su2026sell,teja2026smart}. These differ from product-question answering, fact retrieval, or catalog browsing. Benchmarks such as WebShop and Shopping MMLU evaluate other components of shopping assistance \citep{yao2022webshop,jin2024shopping}; task-and-tool evaluations such as CRMArena and \(\tau\)-bench assess execution rather than rhetorical persuasion \citep{huang2025crmarena,yao2024taubench}. The sales-management literature separately maps the new opportunities and risks of autonomous agents \citep{hunter2026agents,nelson2026inbound}. Work on generative AI in organizational workflows emphasizes that improvements in worker information access cannot automatically be equated to autonomous end-to-end commercial success \citep{brynjolfsson2025work}.

We define five requirements for a successful sales response: (1) respect the buyer's hard constraints; (2) name a justified best-fit option, including no purchase if none fits; (3) answer the buyer's actual objection; (4) fit the length and detail to the buyer's urgency; and (5) propose an action only if its preconditions and capability are real. Repetition of ``would you like'' is not an outcome by itself.

\subsection{What experiments already establish about selling}
Prior controlled studies help locate the present empirical contribution. \citet{fang2026sales} report large-scale online-retail field experiments that evaluate GenAI in commercial workflows; the intervention and outcomes are different from our first-response language audit. \citet{yang2023profit} analyze the commercial penalty of an agent whose profit-maximizing objective is perceived as single-minded. \citet{winter2026selling} extends the domain to settings in which a seller knows its counterpart is another automated agent. These studies show why the question is not whether an LLM \emph{can} produce persuasive text. The organizational question is whether it can reliably perform a fit-aware, buyer-compatible sales process with real execution rights.

In the adjacent sales benchmark literature, \citet{su2026sell} report performance across 15 mainstream models and carefully distinguish junior-intermediate human baselines from expert seller performance; their multi-turn benchmark does not estimate online conversion rates for the present Aiso corpus. Our analytic focus is narrower: natural buyer-prompt language and grounded next steps, using raw conversation records instead of synthetic dialogue scores.

\subsection{Why alignment can be a feature, not merely friction}
We propose two falsifiable hypotheses. \pnt{H1 (adviser--seller separation):} In general-purpose settings, explicit transaction-specific language is less common than informational recommendation or soft follow-up language. \pnt{H2 (commercial prompting trade-off):} In otherwise identical product-choice tasks, a seller-oriented role instruction increases concrete commercial next steps but may increase unsupported promises or otherwise degrade best-fit recommendations. Neither hypothesis predicts that safety constraints should be disabled.

A third hypothesis motivates subsequent work rather than being tested here: \pnt{H3 (specialist opportunity):} An agent trained on commercial conversations and connected to verified merchant systems can improve qualified decision progress relative to a generic model, while respecting customer welfare constraints. The independent Javo field study is relevant to H3 but is not a test of H1 or H2's causal mechanisms \citep{tannenbaum2026selling}.

\section{Data, definitions and research design}
\subsection{Aiso historical conversation corpus}
A read-only audit of Aiso's connected Supabase project on 21 November 2026 found 725,219 records in \texttt{public.conversations}, dated April 2023 to July 2025, and 139,351 records in a supplementary table. Source provenance overlaps were not deduplicated, so the two totals are not presented as unique user count. The present analysis uses the primary table only. The analytic historical filter includes rows from sources \texttt{wc} and \texttt{wc\_4\_8M}, English language, dataset-assigned \texttt{commercial} or \texttt{transactional} intent, and GPT-family model labels including GPT-3.5, GPT-4, GPT-4o/4.1, and o1. That yields 12,178 rows.

\pnt{Important extraction choice.} The stored \texttt{conversation} text sometimes concatenates unrelated user topics into a single row. Turn counts do not, by themselves, show progress in a continuous purchase process. Accordingly, we analyze only the text following the first \texttt{assistant:} delimiter and before the next \texttt{user:} delimiter. We do not treat later turns as shopping outcomes or claim conversion from these records. A hashed conversation identifier supports deterministic reproduction without printing buyer identities.

\subsection{Direct purchase-request screen}
Within the historical analytic filter, we identify a stricter \emph{purchase-request proxy} if the initial prompt contains an expression such as ``should I buy,'' ``which should I buy,'' ``what should I buy,'' ``I want to buy,'' ``help me choose,'' ``recommend a/an,'' or a ``best \ldots\ for my'' construction. The full regular expression is deposited with the SQL source. The screen returns \(n=279\) nonempty first assistant replies. It favors precision over recall but still misclassifies some cases; these are not manually validated shopping-intent labels.

Prior Aiso work likewise finds that a visible answer rarely provides direct evidence of a completed purchase \citep{tannenbaum2026purchase}; search-work reconstruction has distinct measurement goals and does not by itself report sales \citep{tannenbaum2026search}.

An additional \emph{constraint proxy} flags first prompts containing a price, budget, ``for my,'' ``must have,'' or another selected requirement phrase. This is a lexical measure of supplied context, not a determination that the assistant had \emph{all} facts needed to recommend a particular product.

We predefine four overlapping response flags: \emph{recommendation phrase} (e.g., ``recommend,'' ``choose,'' ``buy the,'' ``best option''), \emph{qualification phrase} (e.g., ``however,'' ``it depends,'' ``keep in mind''), \emph{transaction-specific next step} (e.g., ``book a demo,'' ``start a trial,'' ``add to cart,'' ``request a quote''), and \emph{soft follow-up} (e.g., ``if you'd like,'' ``I can help''). A zero in one column is therefore a zero \emph{for that dictionary}, not an absence of semantically equivalent text. All flags can overlap.

\subsection{2026 Aiso prompt-monitoring data}
\texttt{public.snapshot\_prompt\_responses} contains 80,458 GPT and Gemini monitoring response rows across 617 snapshots. It reflects tracked prompts selected or generated for Aiso customer visibility projects. Unlike the historical corpus, it is not a collection of organic consumer conversations. We use 1,050 nonempty, length-screened GPT-5.4--Gemini response pairs on 383 distinct purchase-oriented prompt strings. Section~\ref{sec:intent} stratifies them by original buyer-intent wording, independent of response content. Pairs share the snapshot and prompt identifiers. Where prompts repeat, our primary descriptive rates weight each distinct prompt equally rather than counting every monitoring run as an independent buyer.

For GPT-5.4 responses, we measure the frequency of narrow action expressions, soft informational follow-ups, and selected qualification markers. The design does not observe a merchant site, conversion event, or a true ability to call a booking tool. We therefore use it as a language-distribution cross-check, not a buyer-conversion experiment.

\subsection{Controlled prospective replay}
On 21 November 2026, we ran a GPT-5-nano model through an authenticated Vercel AI Gateway research build. Eight hypothetical product briefs were constructed from \emph{topic categories} observed in Aiso's historical prompts, including flexible offices, transcription tools, charging cables, venues, CRM plans, hotels, data enrichment, and coworking for confidential work. All product specifications were invented and internally consistent; they were \emph{not} actual offers from named companies. The same scenario and buyer utterance were sent in two system-instruction arms: a truthful \emph{general-purpose assistant} and a \emph{responsible sales representative} instructed to recommend a fit, address the objection, and propose a specific next step. No shopping, checkout, calendar, or lead-capture tools were provided to the model.

Of 16 attempted completions, 12 succeeded and four returned platform HTTP 429 rate limits. Five scenarios (S01--S05) have complete matched outputs and form the primary paired case audit (10 successful completions). Two additional successful single-arm scenarios are described only as qualitative boundary cases. Outputs were generated once per available cell, without blind human annotation or repeated stochastic seeds. The open-ended output limit was 800 completion tokens with minimal reasoning effort. The two instruction arms use identical product facts; this is an exploratory manipulation of system prompt role, \emph{not} of training, model weights, or the deployed ChatGPT.com product.

\subsection{Estimands and inference}
For the role experiment, let \(S_{ic}\) denote whether the response to scenario \(i\) under role condition \(c\) proposes a concrete commercial step, and let \(F_{ic}\) denote whether its best-fit recommendation is unambiguous and correct. The paired difference is
\begin{equation}
\widehat{\Delta}_S = \frac{1}{N}\sum_{i=1}^N\left(S_{i,\mathrm{seller}} - S_{i,\mathrm{assistant}}\right), \quad N=5.
\end{equation}
With \(N=5\), percentages are descriptive; exact matched-pair tests have very low power. The data cannot estimate sales conversion, revenue uplift, consumer welfare, or the effect of OpenAI's training policy. For historical data, the percentages represent dataset records satisfying explicit text patterns; no population or causal inference is made.

\section{Results: what real commercial questions reveal}
\subsection{Direct requests elicit advice more often than a sales process}
Table~\ref{tab:historical} reports the new historical first-response audit. In the purchase-request proxy subset, 63/279 replies (22.6\%) use one of our selected recommendation phrases; 109/279 (39.1\%) use a qualification marker; 84/279 (30.1\%) qualify without one of those recommendation phrases. We observe \emph{zero} matches to the narrow transaction-specific next-step dictionary in 279 replies. The absence of dictionary matches cannot establish that no response advanced a buying decision.

\begin{table}[tb]\centering\small
\caption{First-response markers for the 279 strict purchase-request proxy rows in Aiso's historical GPT-family corpus. Categories overlap and are based on exact lexical rules, not human judgments of response quality.}
\label{tab:historical}
\begin{tabular}{lrrrrr}
\toprule
Model family & Rows & Rec. phrase & Qual. phrase & Qual., no rec. & Transaction phrase\\
\midrule
GPT-3.5 & 124 & 30 & 44 & 33 & 0\\
GPT-4 & 70 & 16 & 36 & 28 & 0\\
GPT-4o/4.1 & 63 & 13 & 19 & 15 & 0\\
o1 & 22 & 4 & 10 & 8 & 0\\
\midrule
\textbf{All screened} & \textbf{279} & \textbf{63} & \textbf{109} & \textbf{84} & \textbf{0}\\
\bottomrule
\end{tabular}
\end{table}

These results are consistent with the interpretation that ordinary assistant conversations are not naturally structured as sales funnels. The data do \emph{not} establish that these models were incapable of selling. The records are historical, the intent labels are imperfect, the users were not necessarily visiting a merchant, and a recommendation need not contain one of our lexical markers.

\subsection{Context does not automatically create sales language}
Among 71 strict purchase-request rows that also contain a selected constraint phrase, 15 responses use our recommendation dictionary, 37 contain a qualification marker, and none match the narrow transaction dictionary. This is an especially tentative subgroup: a phrase like ``for my'' does not guarantee enough information for a correct recommendation, and many qualified answers may be appropriate. Nonetheless, the result offers a more targeted motivation for testing decision closure when the user supplies hard preferences. It does not imply that every qualified reply is commercially weak.

\subsection{Recent model outputs remain oriented toward information}
The 2026 monitoring cross-check finds a prompt-weighted 1.11\% prevalence of specific buy/book/demo/trial phrases in GPT-5.4 answers, compared with 4.31\% for soft follow-up offers. A selected qualification-marker dictionary matches 31.24\% of prompt-weighted answers. Although the hard-action and soft-follow-up dictionaries are not exhaustive, the rank ordering is consistent with an information-oriented response style in this particular monitoring workload. It is not evidence of completed sales, and the monitoring prompts differ substantially from the historical user prompts.

\begin{figure}[tb]\centering\includegraphics[width=.84\linewidth]{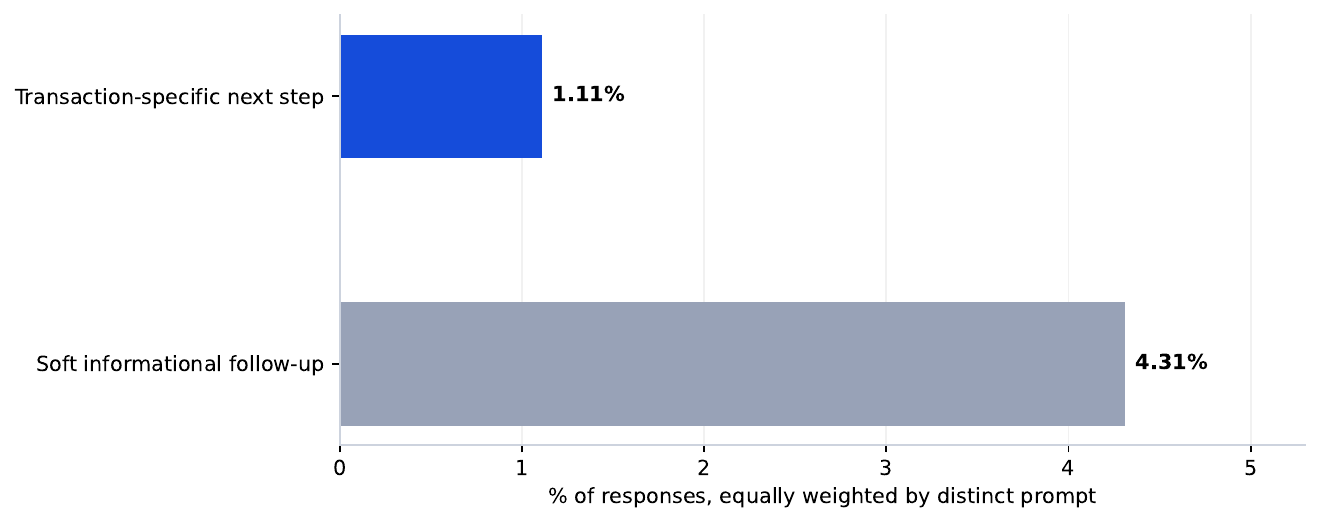}
\caption{Narrow action language versus soft follow-up language in recent GPT-5.4 brand-monitoring answers, weighted equally over 383 distinct purchase-oriented prompts. A lexical screen is not a measure of sales success.}
\label{fig:language}
\end{figure}

\subsection{An explicit buying-intent sensitivity test}\label{sec:strict}
The purchase-request screen in Table~\ref{tab:historical} includes expressions such as ``recommend a,'' which can ask for books, films, or topics rather than a purchase. To test how this affects our descriptive results, we applied an additional \emph{explicit-buying} filter. It requires unambiguous first-person purchase verbs or a choice between items, such as ``I need to buy,'' ``where can I buy,'' ``should I purchase,'' or ``I can't decide between.'' This yields 56 first-response rows from the same historical base. It is a smaller, face-valid subset of the 279-row screen, not an independent dataset.

In those 56 rows, the expanded recommendation-language dictionary matches 24 replies (42.9\%, exact 95\% binomial interval 29.7--56.8\%), a narrower qualification dictionary matches 32 (57.1\%, interval 43.2--70.3\%), and the selected transaction-action dictionary matches none (0/56; exact upper 95\% limit 6.4\%). The constraint proxy is positive in 31 of 56 questions; among those 31, 12 replies have a recommendation phrase and 20 have a qualification marker. We do not equate lexical constraints with enough context for a product to be recommended.

\begin{figure}[tb]\centering\includegraphics[width=.98\linewidth]{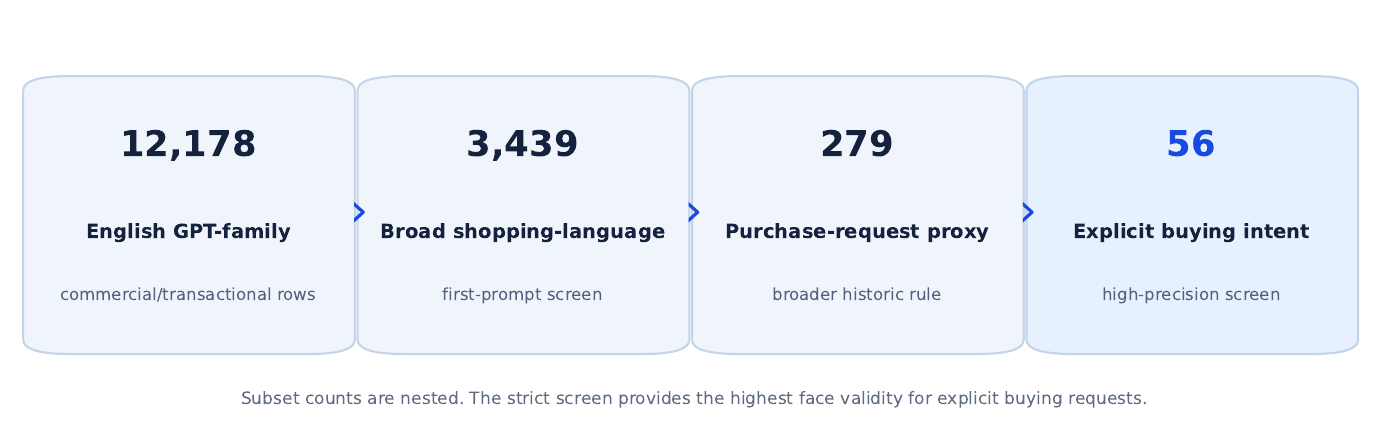}
\caption{Nested historical selection pipeline. A more explicit buying-intent screen is used to check whether recommendations and transactional action language depend on the original permissive screening rule.}
\label{fig:sample}
\end{figure}

\begin{figure}[tb]\centering\includegraphics[width=.97\linewidth]{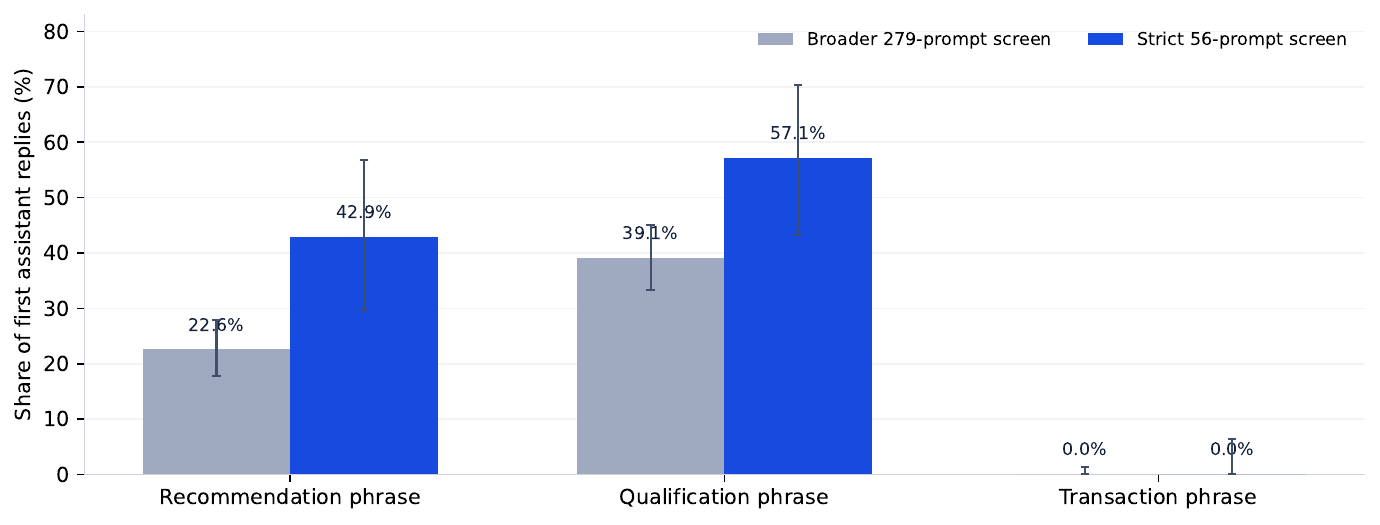}
\caption{Observed first-response language under the original 279-row screen and the narrower 56-row explicit-buying screen. Error bars are exact 95\% binomial intervals for the descriptive row fractions. The coding dictionaries differ slightly and are frozen in the two supplied SQL files; figure rates should not be used as a model-accuracy score.}
\label{fig:sensitivity}
\end{figure}

The two screens serve different purposes. The 279-row filter captures more of the recommendation ecosystem, whereas the 56-row filter targets buyers expressing the act of purchase. The comparable result across both is that the first assistant answer rarely includes a merchant-specific next step, even when it discusses useful products and tradeoffs. This pattern is consistent with advisory behavior rather than a merchant checkout dialogue.

\subsection{Buy, recommend or compare: the strongest 2026 distinction}\label{sec:intent}
We further divide the 383 distinct 2026 monitored prompts using the prompt text alone, before viewing the answer markers. The deterministic hierarchy assigns a prompt to \emph{transaction wording} if it contains ``buy,'' ``book,'' ``purchase,'' ``order,'' ``checkout,'' ``sign up,'' or related terms; otherwise to \emph{recommendation wording} if it asks which option, asks for a best choice, or requests a recommendation; otherwise to \emph{price/comparison wording}. The resulting mutually exclusive groups have 42, 221 and 120 unique prompts, respectively.

\begin{table}[tb]\centering\small
\caption{Transaction-specific language by buyer-prompt intent in the 2026 monitored GPT-5.4 and Gemini answer pairs. Equal weight per distinct prompt, averaging repeated runs. SE is calculated across distinct prompts. All numbers are newly recomputed from Aiso's read-only monitoring tables.}
\label{tab:intent}
\begin{tabular}{lrrrrrr}
\toprule
Prompt segment & Prompts & Pairs & GPT hard action & Gemini hard action & GPT SE & Gemini SE\\
\midrule
Price / comparison & 120 & 336 & 0.21\% & 1.04\% & 0.21 pp & 0.86 pp\\
Recommendation & 221 & 615 & 0.00\% & 0.00\% & 0.00 pp & 0.00 pp\\
Transaction wording & 42 & 99 & 9.52\% & 14.29\% & 3.71 pp & 5.12 pp\\
\bottomrule
\end{tabular}
\end{table}

\begin{figure}[tb]\centering\includegraphics[width=.92\linewidth]{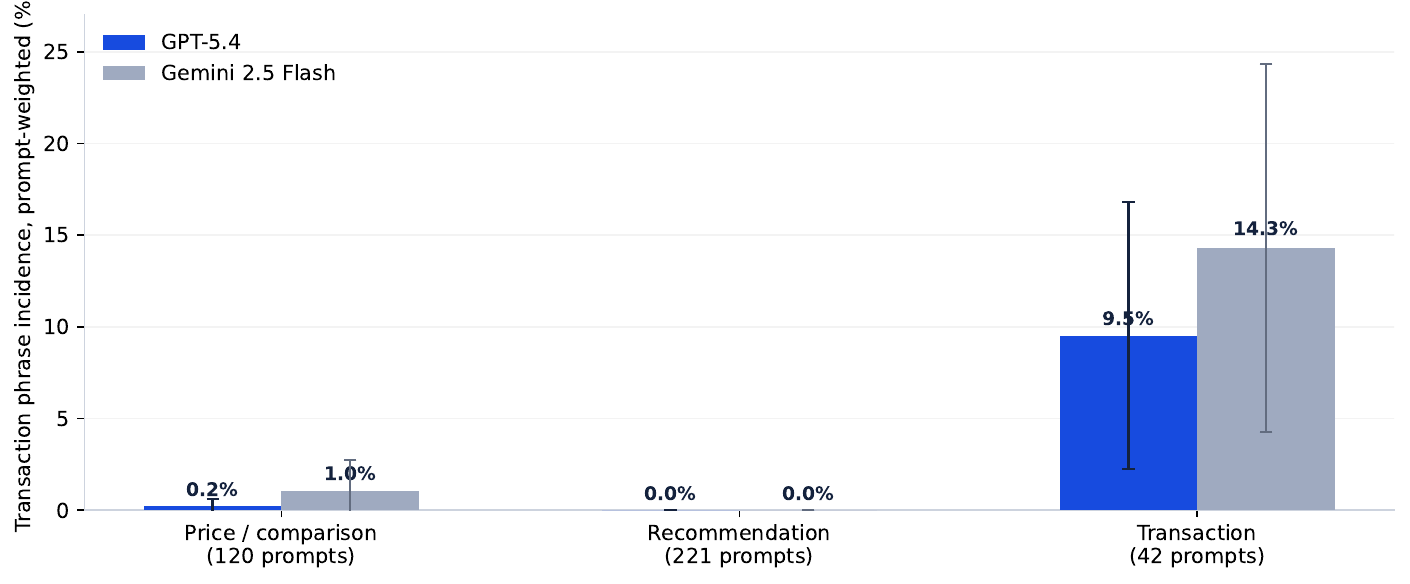}
\caption{A buyer's explicit intent changes how often a general-purpose model uses concrete transaction phrases. Both models show the same qualitative intent gradient. Error bars represent $\pm1.96$ prompt-level SE and are not constrained to zero at the lower bound.}
\label{fig:intent}
\end{figure}

The strongest descriptive contrast is between prompts that explicitly mention a transaction and prompts that merely ask for a recommendation. Seven of 42 transaction-worded prompts elicited at least one GPT hard-action marker across their repeated answers; zero of 221 recommendation-worded prompts did. Fisher's exact test on the \emph{prompt-level ever-action} indicators gives $p=1.69\times10^{-6}$ (one-sided); this is a contrast of prompt categories in a selected monitoring workload, not the causal effect of altering a prompt. The parallel Gemini contrast is also seven of 42 versus zero of 221. The entire comparison uses 1,050 matched answer pairs and retains distinct prompts as units for summary inference.

A distinct commercial point follows from these measurements: the general assistant is capable of transaction-related phrasing when the customer explicitly asks in transactional terms, but its default language in comparison and recommendation questions is far less likely to advance an authorized purchase. A specialist sales system can reasonably detect \emph{qualified intent} and offer the next step proactively rather than requiring the buyer to supply that step in the question.

\section{Results: the sales-role intervention}
\subsection{Commercial prompts change closing behavior}
The five complete paired sales scenarios show an increase from 2/5 to 5/5 in concrete next-step proposals under seller instructions. The estimated paired difference is \(+3/5=60\) percentage points, but with three discordant pairs and only five scenarios, an exact two-sided McNemar test yields \(p=0.25\). The result is a small demonstration of prompt sensitivity, not statistically reliable evidence of a population uplift. The extra language includes starting a trial, beginning a reservation, providing a quote, or shipping an item.

\subsection{The same instruction does not assure product fit}
An unambiguous, non-contradictory best-fit product selection appears in 3/5 replies in \emph{each} condition. In the flexible-office case, the buyer explicitly says they may outgrow their space after six months. Product A has a six-month term; Product B is a 24-month contract with a lower monthly price. The seller-style completion eventually recommends B, privileging the price discount over the buyer's uncertainty. The general reply is indecisive, but it does not insist on the 24-month term. Stronger sales language can therefore accompany a worse fit.

In the seven-person event case, both replies initially name the cheaper six-seat venue as the better fit before correctly concluding that the eight-seat venue is required. This is a failure of answer coherence and headline choice, not of arithmetic. A buyer skimming the first sentence might make the wrong selection even when the full response later repairs the error.

\begin{table}[tb]\centering\small\caption{Manual audit of five successfully paired prospective scenarios. Synthetic catalog facts, real model completions. ``Action'' indicates concrete commercial language, not an executed action.}
\label{tab:pilot}
\begin{tabularx}{\linewidth}{p{1.35cm}p{2.3cm}XX}
\toprule
Case & Buyer priority & General assistant & Sales-role instruction \\
\midrule
S01 Office & Six-month uncertainty & Describes conditional trade-off; no clear choice & Recommends 24-month term; offers to lock in\\
S02 Transcription & 90 hours, SRT, cost & Selects Plan A; offers usage analysis & Selects A; proposes trial, adds unverified entitlements\\
S03 Cable & Durability under budget & Correctly selects B; no order step & Selects B; offers shipment with no tool\\
S04 Venue & Seven seats, no list & Starts with undersized A, then says B & Also starts with A; offers booking hold\\
S05 CRM & SSO mandatory & Correctly selects A; suggests account initiation & Selects A; offers quote and setup call\\
\bottomrule
\end{tabularx}
\end{table}

\begin{figure}[tb]\centering\includegraphics[width=.91\linewidth]{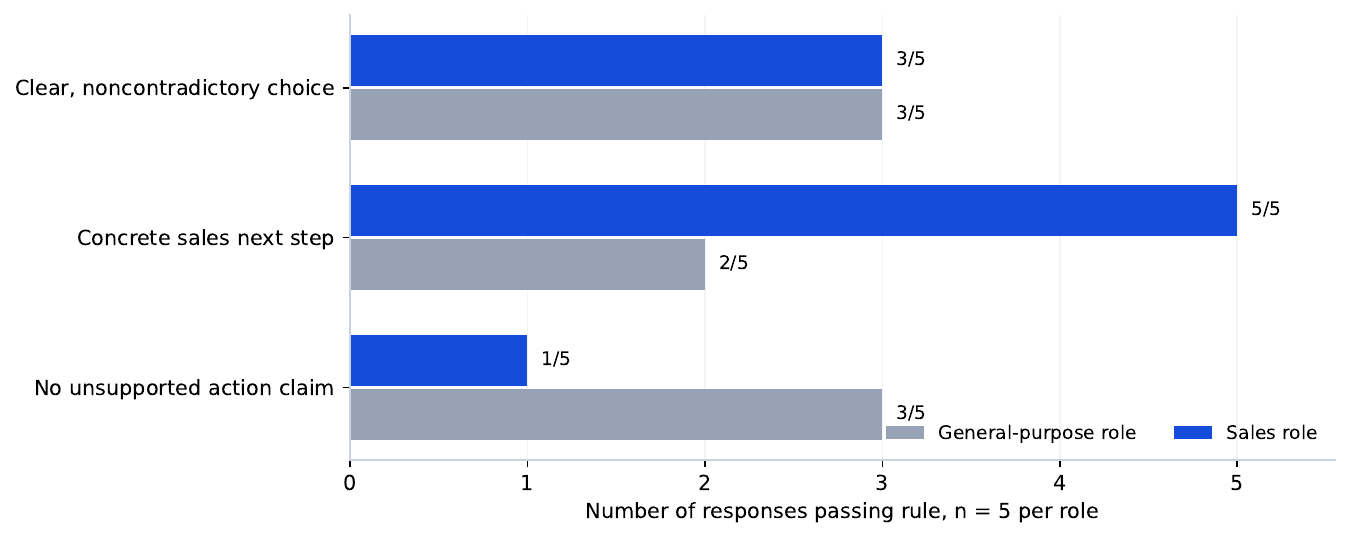}
\caption{Exploratory audit of the five matched cases, using transparent hand-coded rules. Results are descriptive and should not be interpreted as a benchmark score for deployed ChatGPT.}
\label{fig:pilot}
\end{figure}

\subsection{A gap between persuasive promises and real tools}
Four of five seller-style completions, versus two of five general-assistant completions, offered an operational action unsupported by the experiment's tool inventory. The seller suggested shipping a cable, holding a venue, locking an office, or arranging a service call. These were \emph{offers}, not actual transactions, but in a business context they imply a capability. The sales instruction increased the likelihood of this behavior without installing checkout, booking, CRM, or lead-capture tools. This is a particular risk for an organization that evaluates only the presence of closing language.

One unpaired boundary case concerns a buyer seeking confidential, soundproof space for therapy. The only available fictitious product was a shared desk with meeting-room access. The seller-styled reply acknowledged the mismatch but still discussed enrolling in the unsuitable plan while waiting for a private office. Because the comparison cell was rate-limited, this case is excluded from paired counts; it illustrates why ``no sale'' must remain a valid and sometimes optimal outcome for a responsible seller.

\subsection{Long answers when buyers ask for a decision}
The general-purpose replies average 253 words across matched S01--S05 and the seller replies average 268. In the urgent office inquiry the two replies contain 427 and 473 words, respectively; in the event case, despite the request not to receive another list of pros and cons, they contain 200 and 236 words. Sales-role instruction improved action wording without reliably compressing the answer. The commercial effect of this excess length was not experimentally measured.

\begin{figure}[tb]\centering\includegraphics[width=.94\linewidth]{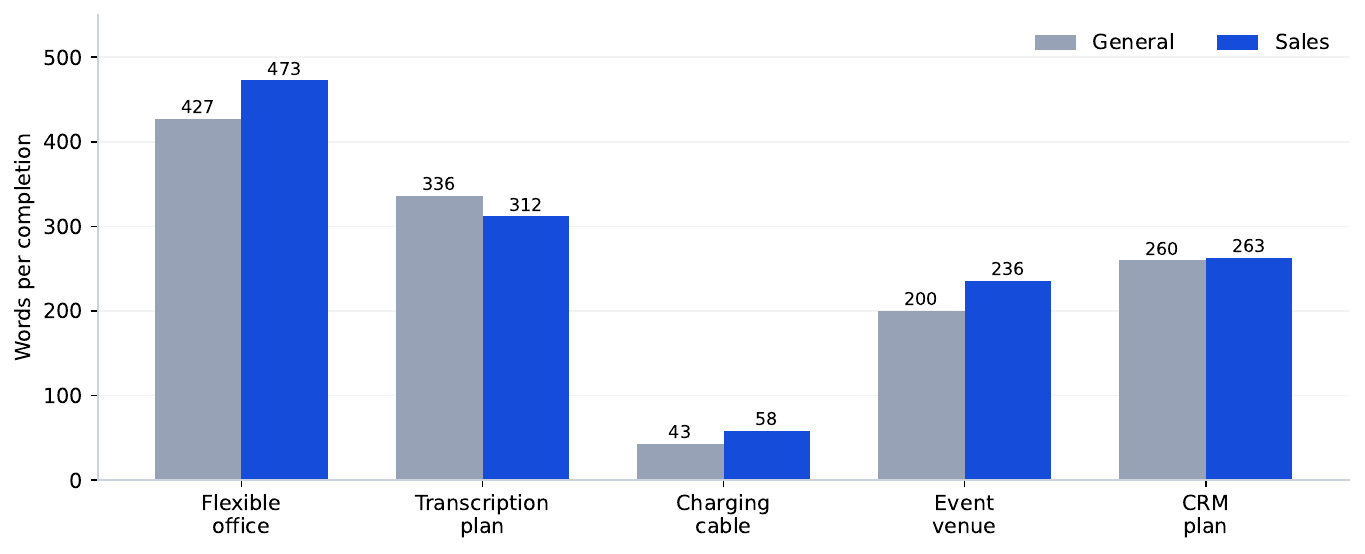}
\caption{Response lengths in the five matched GPT-5-nano scenario pairs. Explicit requests for brevity were present in some buyer statements.}
\label{fig:length}
\end{figure}

\subsection{What the intervention does, and does not, establish}
The role experiment offers evidence that \emph{system prompting alone} can change the surface of commercial behavior while leaving substantive fit and action grounding unresolved. Because the sample is small, synthetic, and non-blinded, we neither claim that ChatGPT.com behaves identically nor quantify an average impact on conversion. GPT-5-nano was selected because that model was accessible under the connected research account's tier; an attempted GPT-6/Gemini-3.1 replay was blocked by the provider's account restrictions. The study is not a comprehensive frontier-model benchmark.

\section{The commercial opportunity: what the Javo field evidence does show}
The observed advice-to-action transition motivates a task-specialization opportunity. The direct randomized comparison of Javo against a frontier sales agent appears in a separate field experiment, not in the corpus language measurements here. That comparison belongs to the separately published \emph{Selling in the Wild} study \citep{tannenbaum2026selling}. In that company-authored research, Javo was trained on 4.6 million consented and de-identified commercial conversations and evaluated in a visitor-randomized field design covering 100 websites and 14.8 million visits. The manuscript reports that, among engaged visitors, the trained agent increased conversion by 21.6\% relative to a frontier general-purpose model placed in the same interface, retrieval, tools, and guardrails, with reported 95\% confidence interval \(+16.9\) to \(+26.3\)\%. Relative to no agent, reported revenue per visitor increased 10.0\% (95\% CI \(+5.1\) to \(+14.9\)\%). Its reported safety checks included unsupported product claims, returns, complaints, and cancellations.

These are results \emph{reported by the prior Javo field paper}, not reconstructed or newly validated from the Supabase subsets analyzed here. They are commercially relevant because the agent comparison held its deployment scaffold constant, isolating the specialist model within that study's design. They do not show that \emph{the absence of caveats} caused the uplift, that GPT-6 would perform identically to the earlier frozen frontier model, or that Javo must outperform every current general assistant. This distinction is essential to scientific interpretability.

\begin{table}[tb]\centering\small\caption{Separation of direct observations, prior experimental evidence, and hypotheses.}
\label{tab:evidence}
\begin{tabularx}{\linewidth}{p{2.7cm}X X}
\toprule
Evidence stream & Directly supported claim & Not established\\
\midrule
Aiso historical first replies & Narrow transaction language absent in 279 screened direct-purchase prompts & Absence of helpful advice or observed lost sales\\
Aiso 2026 monitoring & Hard-action phrases rare in selected GPT-5.4 answers & Representative shopper behavior or conversion outcomes\\
Prospective GPT-5-nano replay & Seller role changes next-step language; tool and fit errors observed in cases & Average impact on revenue, modern ChatGPT.com, or training causes\\
Javo randomized field paper & Reports commercial uplift of a specialized agent versus a frontier comparator under shared tools & Caveat-tax mechanism or universal superiority\\
\bottomrule
\end{tabularx}
\end{table}

\section{Discussion: probably a good thing for users, a design problem for merchants}
\subsection{A buyer-aligned assistant should not be a merchant's commission engine}
A general-purpose assistant may encounter a buyer comparing multiple brands with no reason to privilege one merchant's revenue. Displaying alternatives, declining to invent availability, or telling a customer that their requirements cannot be met can preserve agency. It may look like weak sales behavior from the merchant's perspective, but such neutrality is an appropriate property of a user-oriented system. This aligns with OpenAI's stated emphasis on honesty, minimizing harm, and user agency \citep{openai2026spec} and with research on reactions to identifiable profit motives \citep{farmer2026profit}.

A public consumer adviser need not become aggressively promotional for a merchant to benefit from AI-driven commerce. Instead, brands that want to convert qualified traffic need a separate agent with an explicitly commercial role, clear disclosure, access to verified merchant facts, and tools that can actually execute the proposed next step.

\subsection{A specialist sales agent is more than a persuasive prompt}
The replay study offers a practical distinction. A system instruction to sell increased next-step offers but also produced ungrounded commitments. To avoid repeating these failures, a purpose-built agent needs a fit validator that treats user requirements as hard constraints, retrieval of inventory and policy terms, an action authorizer, traceable provenance for claims, and a negative-outcome objective that allows refusal or human handoff when no suitable product exists. Training on real commercial conversations may improve the model's understanding of conversational progression, but no training scheme is sufficient without grounded external state and oversight.

\begin{figure}[tb]\centering\includegraphics[width=.94\linewidth]{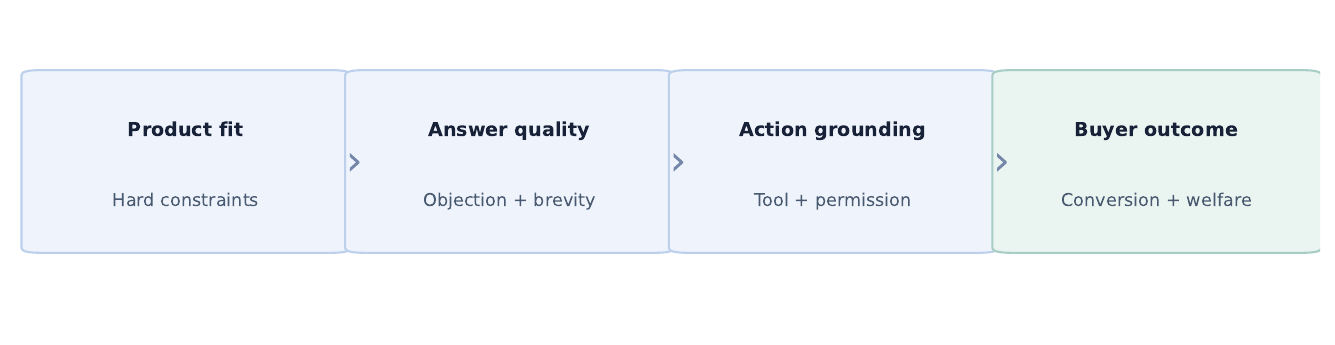}
\caption{Proposed evaluation stack for a production merchant agent. All four layers must be evaluated independently. A conversion metric alone cannot certify a trustworthy salesperson.}
\label{fig:pipeline}
\end{figure}

\subsection{What should be optimized?}
Let a merchant agent optimize a constrained objective
\begin{equation}
\max_\pi\; \mathbb{E}[\mathrm{QualifiedProgress}(\pi)] - \lambda\,\mathbb{E}[\mathrm{ActionMisrepresentation}(\pi)]
\end{equation}
subject to \(\mathrm{ProductFit}\geq \tau_f\), \(\mathrm{ClaimSupport}\geq \tau_c\), and \(\mathrm{ConsumerHarm}\leq \tau_h\). This is a normative design proposal, not a fitted model. A referral, booking, or lead should count positively only if it is qualified and the system is authorized to perform it. Returning ``none of our plans meets your requirement'' is a correct decision in many cases. A future field trial should register both merchant outcomes and consumer guardrails, not conversion alone.

\section{Scope, reproducibility and disclosure}\label{sec:ethics}
The historical and monitoring samples differ in provenance: the former stores user conversations from 2023--2025, the latter records repeated brand-tracking prompt responses in 2026. All corpus measures are predeclared string matches and are not conversion labels. The controlled test uses fictional catalog facts and GPT-5-nano API outputs, with five matched scenarios. Its incomplete cells (HTTP 429) are not scored. SQL, prompt dictionaries, source manifests, manual coding definitions and aggregate data accompany the paper. No identifiable historical conversation text is reproduced.

Aiso Research, AISO Boost Ltd. is commercially associated with Javo. The author cites a separate Javo field study, and its results are attributed to that paper rather than combined statistically with this manuscript's observations. Reproducing the field effect independently and determining its mechanisms remain distinct experimental tasks.

\section{Conclusion}
General-purpose assistants often act as buyer-aligned advisers rather than merchant sales representatives. In Aiso's historical corpus, stricter direct-purchase screens still produce almost no transaction-specific first-response language. In 2026 model monitoring, GPT and Gemini show a pronounced step-change in that language only when the buyer uses transactional wording. Sales-role prompting increases closing language in a controlled pilot, but does not reliably fix product fit or action grounding. The paper does not prove that GPT is incapable of selling, or that safety-oriented training caused the observed differences. It establishes the sharper product distinction: \emph{advising a user and acting as a merchant's authorized salesperson are different tasks}. Separate Javo field evidence reports a conversion advantage for commercial-conversation training in a controlled deployment. The promising next step is to combine learned sales skill with verifiable product fit and executable, bounded actions, while preserving the buyer's ability to walk away.

\section*{Data availability and reproducibility}
Aggregate counts, filters, SQL scripts, the annotated synthetic replay manifest, figure-generation code, and the complete \LaTeX{} source accompany this manuscript in the research submission package. The historical conversation bodies are not redistributed because provenance, privacy and consent conditions vary by source; the Supabase analytics rely on authorized access to Aiso's environment. API replay outputs are linked in the source manifest and summarized in deposited annotations. The independent Javo study is available as arXiv:2610.25461. All figures are vector PDFs. The PDF was compiled with \texttt{pdflatex} and \texttt{bibtex}.

\section*{Acknowledgments}
Aiso Research is responsible for the analysis, interpretation, and preparation of this manuscript. No user-facing conversion or purchasing actions were undertaken in the experiments reported here.

\bibliographystyle{plainnat}
\bibliography{references}

\appendix
\section{Historical regex protocol and exact denominators}
For reproducibility, \texttt{sql/historical\_purchase\_screen.sql} contains a read-only query with the actual source and model filters and response-marker dictionaries. Rows are filtered to English, labeled commercial or transactional, the two historical \texttt{wc} sources, and GPT-family model tags. We then take the first \texttt{assistant:} segment before the next \texttt{user:} segment. This avoids treating concatenated unrelated topics as a continuous shopper journey. The strict purchase-request screen is applied to the first prompt. The observed subgroup counts are shown in Table~\ref{tab:history_appendix}.

\begin{table}[h]\centering\small\caption{Strict direct-purchase screen, by a lexical context-constraint proxy. ``Qual., no rec.'' means qualifier marker present but selected recommendation marker absent.}
\label{tab:history_appendix}
\begin{tabular}{lrrrrr}\toprule
Family, constraint & \(n\) & Rec. & Qual. & Qual., no rec. & Hard next step\\\midrule
GPT-3.5, no & 92 & 23 & 29 & 22 & 0\\
GPT-3.5, yes & 32 & 7 & 15 & 11 & 0\\
GPT-4, no & 50 & 11 & 21 & 17 & 0\\
GPT-4, yes & 20 & 5 & 15 & 11 & 0\\
GPT-4o/4.1, no & 50 & 11 & 15 & 12 & 0\\
GPT-4o/4.1, yes & 13 & 2 & 4 & 3 & 0\\
o1, no & 16 & 3 & 7 & 5 & 0\\
o1, yes & 6 & 1 & 3 & 3 & 0\\
\bottomrule\end{tabular}
\end{table}

\section{Replay codebook and scenario specifications}
The five matched scenarios are listed in the research package's \texttt{data/matched\_pilot\_annotations.csv}. \emph{Clean best-fit} requires a noncontradictory selected product consistent with the brief's hard constraints. \emph{Concrete next step} requires an explicit trial, order, booking, quote, or similar forward action; additional comparative advice does not qualify. \emph{Unsupported action} flags a claim that the agent can actually initiate shipment, hold, quote, or booking when no such tool exists; an invitation for the user to visit the merchant is not automatically disallowed. Responses were reviewed with the same coding definitions in both arms and can be reannotated independently using the linked raw-text manifest.

The pilot is exploratory; the labels are not inter-annotator agreement estimates. Some coding is arguable, notably whether ``I can initiate an account'' constitutes an ungrounded offer rather than a helpful invitation, and whether initially identifying a bad option followed by correction should be credited as a clean recommendation. The source package reports the actual text and the decision rule so readers can apply alternative reasonable codings.

\section{Exact tests, unit of analysis and lexical sensitivity}\label{app:tests}
For a binary historical marker observed $k$ times among $n$ screened records, the displayed descriptive proportion is $k/n$ and its 95\% interval is the exact Clopper--Pearson interval. For the zero-of-56 transaction marker, the upper two-sided 95\% bound is 6.38\%. We do not treat overlapping regex flags as independent or combine the 56-row subset with its parent 279-row sample. The agreement of the zero observed hard-action counts is a robustness check across operational definitions.

For the monitored prompt cohort, let $y_{prm}$ equal one if response $r$ to distinct prompt $p$ under model $m$ matches the selected dictionary. Define the prompt mean $\overline y_{pm} = R_p^{-1}\sum_r y_{prm}$, and report $P^{-1}\sum_p\overline y_{pm}$ with a standard error estimated from the distribution of prompt means, rather than treating $\sum_pR_p$ repeated answers as independent buyers. The category comparison of \emph{at least one action marker across replays} uses a two-sided $2\times2$ table at the prompt level; the directional Fisher $p$-value is reported with its direction. This test is not a randomized intervention on prompt wording.

The regex-based method allows exact database-scale replication but has a different interpretation from semantic human coding. The new explicit-buying screen excludes generic ``recommend a'' and ``best'' queries that could refer to non-purchasable content. The primary first-response analysis uses only the text between the first assistant and next user delimiter to avoid interpreting concatenated topics as one ongoing sales session. All model comparisons are descriptive, not controls for differences in user populations or request dates.

\section{Commercial agent evaluation rubric}\label{app:rubric}
The next experimentally testable quality metric is \emph{qualified action readiness}: a product is recommended only when its documented attributes satisfy all buyer hard constraints; the reply answers the buyer's expressed objection in an appropriate amount of text; a genuine next step exists in the current merchant's tool inventory; and each claimed action can be replayed against a confirmed tool outcome. A response can pass recommendation but fail execution. An agent can pass consumer welfare by declining to sell when no item qualifies.

We recommend a seven-label annotation instrument: (1) buyer context complete; (2) fit satisfied; (3) initial recommendation unambiguous; (4) objection addressed; (5) response proportional in length; (6) proposed next step authorized; (7) action completed and verified. Label (7) is unavailable in the observational Aiso datasets; no rate for it is reported. The field Javo experiment supplies merchant outcomes under its own design, and should not be reinterpreted as measurements of these seven labels in the historical GPT corpus.

\section{Field experiment required to test the mechanism}
To test whether buyer-first caution itself changes commercial outcomes, a future preregistered study should randomize actual consenting merchant visitors among: (A) a buyer-oriented general agent with verified catalog and tools; (B) the same agent and tool scaffold with a merchant-sales system instruction; and (C) a specialist trained sales model under identical tools and guardrails. The primary unit should be the visitor, not the generated utterance; assignment should be stable and stratified by site. Primary outcomes should include qualified purchase conversion and revenue per visitor; guardrails should include returns, complaints, unsupported claims, total cost disclosure, missed-fit rate, and explicit human handoff. Human coding should be blind to arm and assess advice accuracy, buyer-welfare protection, and truthful action capability. This would test the distinct effects of role prompting, model specialization and authorized tools. The current paper does not report such an experiment.

\end{document}